\documentclass{easychair}

\usepackage{microtype}
\usepackage{mathtools}
\usepackage{cite}
\usepackage{latexsym,amssymb,amsmath}
\usepackage{textcomp}
\usepackage{bold-extra}
\usepackage[inline]{enumitem}
\usepackage{xspace}
\IfFileExists{doi.sty}{\RequirePackage{doi}}
{\RequirePackage{hyperref}\def\doi##1{##1}}
\usepackage{xcolor}
\definecolor{linkblue}{RGB}{0,0,180}
\usepackage{hyperref}
\hypersetup{%
  pdftex,%
  colorlinks=true,%
  pdfborder={0 0 0},%
  citecolor=linkblue,%
  urlcolor=linkblue,%
  filecolor=linkblue,%
  linkcolor=linkblue%
}

\newcommand{\ACP}{\textsf{ACP}\xspace}

\newcommand{\ACPH}{\textsf{ACPH}\xspace}
\newcommand{\SAIGAWA}{\textsf{Saigawa}\xspace}
\newcommand{\COCOCO}{\textsf{CO3}\xspace}
\newcommand{\CONCON}{\textsf{ConCon}\xspace}

\newcommand{\CSI}{\textsf{CSI}\xspace}
\newcommand{\CSIho}{\textsf{CSI}$\mathbf{\hat{~}}$\textsf{ho}\xspace}

\title{CoCoWeb \\[.5ex] \Large
A Convenient Web Interface for Confluence Tools\thanks{This work is
supported by the Austrian Science Fund (FWF): project P27528.}}

\author{Julian Nagele \and Aart Middeldorp}

\institute{
  Department of Computer Science,
  University of Innsbruck, Austria \\
  \email{\{julian.nagele,aart.middeldorp\}@uibk.ac.at}
}

\authorrunning{Nagele and Middeldorp}
\titlerunning{CoCoWeb}

\begin{document}

\maketitle

\begin{abstract}
We present a useful web interface for tools that participate in the annual
confluence competition.
\end{abstract}

\section{Introduction}

In recent years several tools have been developed to automatically prove
confluence and related properties of a variety of rewrite formats.
These tools compete annually in the confluence competition~\cite{AHNNZ15}
\href{http://coco.nue.riec.tohoku.ac.jp/}{(CoCo)}.%
\footnote{\url{http://coco.nue.riec.tohoku.ac.jp}}
Most of the tools can be downloaded, installed, and run on one's local
machine, but this can be a painful process.%
\footnote{StarExec provides a VM Image with their environment,
which can helpful in case a local setup is essential.}
Few confluence tools---we are
aware of \COCOCO~\cite{NKYG15}, \CONCON~\cite{SM14}, and
\CSI~\cite{NFM17,ZFM11b}---provide a convenient web interface to painlessly
test the status of a system that is provided by the user.

Inspired by the latest web interface of \CSI~\cite{NFM17},
in this note we present \textsf{CoCoWeb}, a web interface
that provides a single entry point to all tools that participate in CoCo.
\textsf{CoCoWeb} is available at
\begin{center}
\url{http://cl-informatik.uibk.ac.at/software/cocoweb}
\end{center}
The typical use of \textsf{CoCoWeb} will be to test whether a given
confluence problem is known to be confluent or not. This is useful when
preparing or reviewing an article, preparing or correcting exams about
term rewriting, and when contemplating to submit a challenging problem
to the confluence problems
(\href{http://cops.uibk.ac.at/}{Cops})%
\footnote{\url{http://cops.uibk.ac.at}}
database. In particular,
\textsf{CoCoWeb} is useful is when looking for
(killer) examples to illustrate a new technique. For instance, in
\cite{IOS16} a rewrite system is presented that can be shown to be
confluent with the technique introduced in that paper. The authors write
``Note that we have tried to show confluence [\,\dots] by confluence checker
\ACP and \SAIGAWA, and both of them failed.'' Despite having an easy to use
web interface, \CSI was not tried. \textsf{CoCoWeb} could also be
useful for the CoCo steering committee when integrating newly submitted
problems into Cops.

The tools run on the same hardware, which is compatible with a single
node of StarExec~\cite{SST14} that is used for CoCo, allowing for a proper
comparison of tools.

In the next section we present the web interface of \textsf{CoCoWeb}
by means of a number of screenshots. Implementation details are
presented in Section~\ref{sec:implementation} and in
Section~\ref{sec:extensions} we list some possibilities for
extending the functionality of \textsf{CoCoWeb} in the future.

\section{Web Interface}
\label{sec:web interface}

\begin{figure}[t]
\centering
\includegraphics[width=\textwidth]{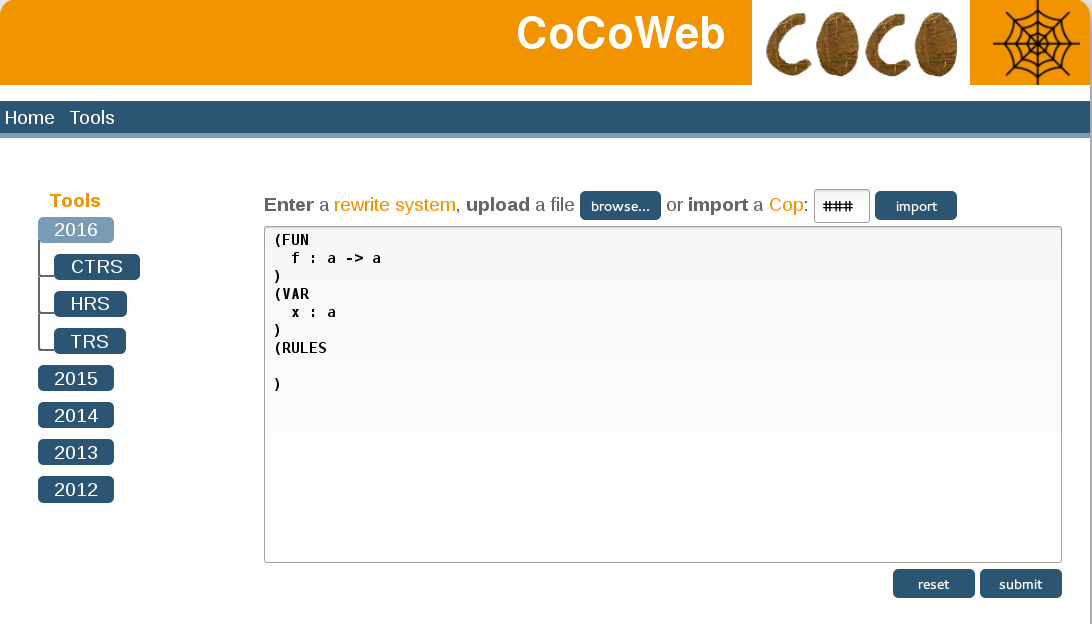}
\caption{The entry page of \textsf{CoCoWeb}.}
\label{fig:1}
\end{figure}

Figure~\ref{fig:1} shows a screenshot of the entry page of
\textsf{CoCoWeb}. Problems can be entered in three
different ways:
\begin{enumerate*}[label=(\arabic*)]
\item using the text box,
\item uploading a file,
\item entering the number of a system in Cops.
\end{enumerate*}
The tools that should be executed can be selected from the tools panel
on the left.
\begin{figure}[b]
\centering
\includegraphics[width=\textwidth]{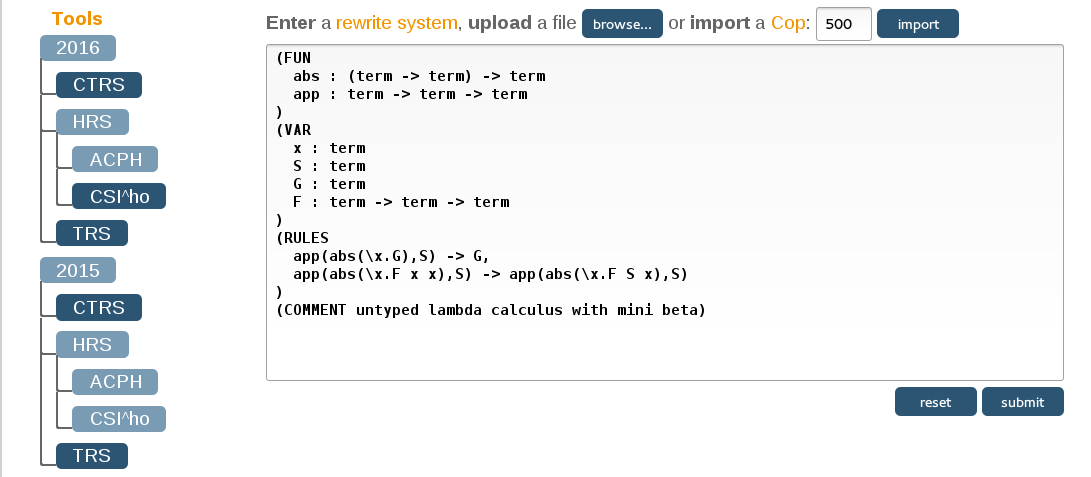}
\caption{Problem and tool selection in \textsf{CoCoWeb}.}
\label{fig:2}
\end{figure}
Tools are grouped into categories, similar to the grouping
in CoCo except that we merged the certified categories with the
corresponding uncertified categories. Multiple tools can be selected.
This is illustrated in Figure~\ref{fig:2}.
Here we selected the CoCo 2016
and 2015 versions of \ACPH and the CoCo 2015 version of \CSIho, and
Cop 500 is chosen is input problem.

\begin{figure}[b]
\centering
\includegraphics[width=\textwidth]{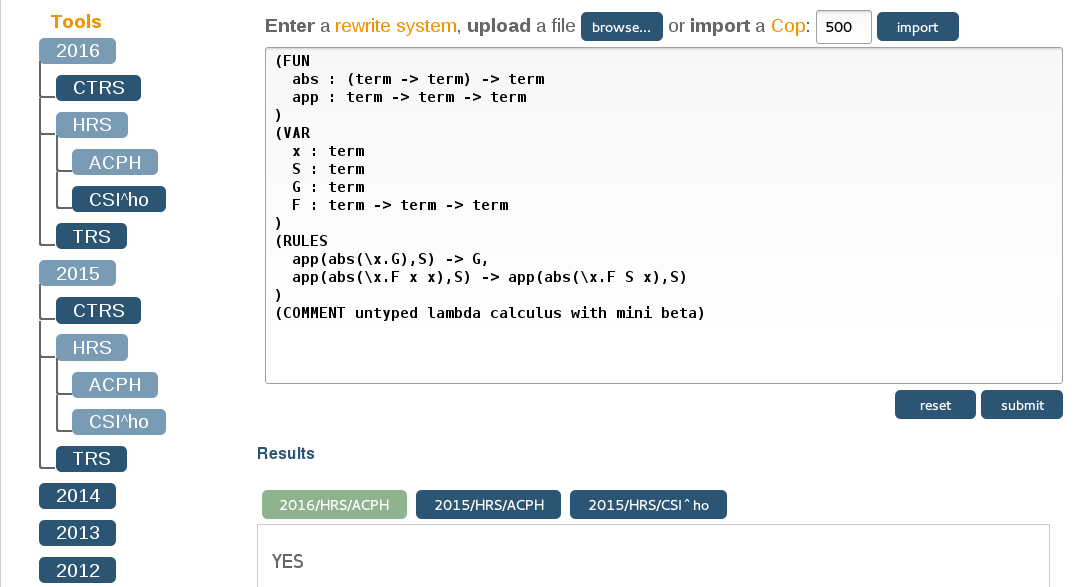}
\caption{Result displaying in \textsf{CoCoWeb}.}
\label{fig:3}
\end{figure}

The screenshot in Figure~\ref{fig:3} shows the output of \textsf{CoCoWeb}
after clicking the submit button. The output of the selected tools is
presented in separate tabs. The colors of these tabs reveal
useful information: Green means that the tool answered yes,
red (not shown) means that the tool answered no, and a maybe answer or
a timeout is shown in blue. By clicking on a tab, the color is made
lighter and the output of the tool is presented. The final line of the
output is timing information provided by \textsf{CoCoWeb}.

\begin{figure}[tb]
\centering
\includegraphics[width=\textwidth]{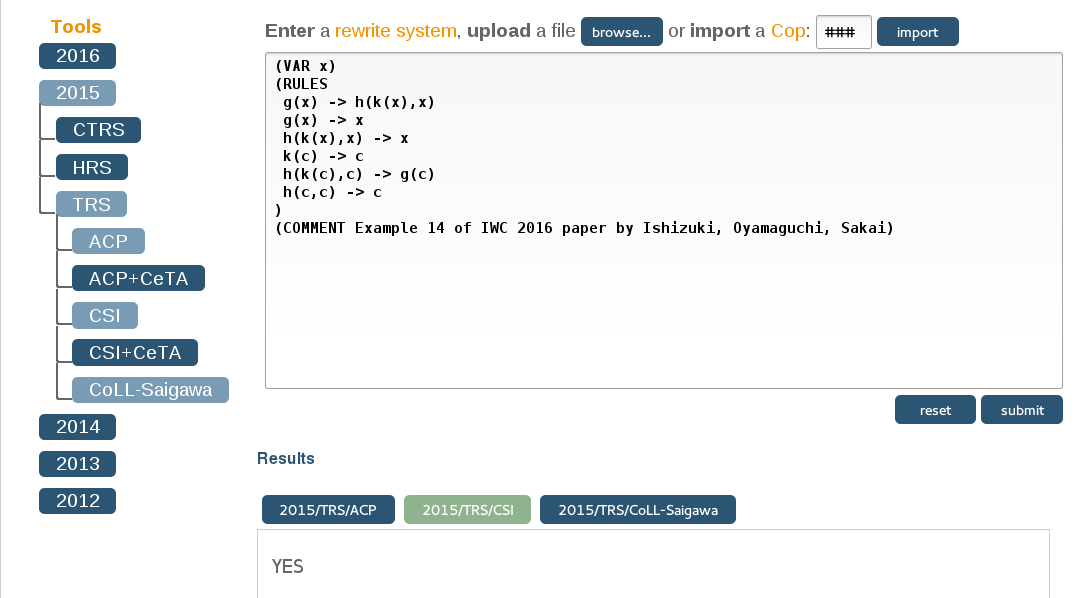}
\caption{Testing Example~14 from \cite{IOS16} in \textsf{CoCoWeb}.}
\label{fig:4}
\end{figure}

The final screenshot (in Figure~\ref{fig:4}) is concerned with the example
in \cite{IOS16} that we referred to in the introduction.

\section{Implementation}
\label{sec:implementation}

Most of \textsf{CoCoWeb} is built using PHP. User input in forms, i.e.,
rewrite systems and tool selections, is sent using the HTTP POST method.
The dynamic parts of the website, namely folding and unfolding in the tool
selection menu and the tabs used for viewing tool output are implemented
using JavaScript.

To layout the tool selection menu we made extensive use of CSS3 selectors.
For instance, the buttons to select tools are implemented as checkboxes
with labels that are styled according to whether the checkbox is ticked or
not:
\begin{verbatim}
        .tools input[type="checkbox"]:checked + label {
          color: white;
          background-color: #799BB3;
        }
\end{verbatim}
Drawing the edges of the tree menu is also done using CSS, relying mainly
on its \texttt{::before} selector.

The content of the tool menu, i.e., years, the grouping by categories, and
the actual tools, is generated automatically from a directory tree that
has the structure of the menu in \textsf{CoCoWeb}.
There small configuration files reside that specify how the tools are
to be run, in case they are selected. Two environment variables are set in
such a file, for example the one for the 2012 version of \SAIGAWA reads
as follows:
\begin{verbatim}
        TOOLDIR="Saigawa-2012/bin"
        TOOL="./starexec_run_saigawa -t $TO $FILE"
\end{verbatim}
The variable \texttt{TOOLDIR} specifies the directory that contains the
tool binary, while \texttt{TOOL} gives the tool invocation, which in
turn refers to \texttt{TO}, the timeout, and \texttt{FILE}, the input
rewrite system. Using such configuration files tools are run using the
following script, whose first and second argument are the configuration
of the tool and input rewrite system respectively:
\begin{verbatim}
        DIR=$(pwd -P)
        FILE=$(readlink -f $2)
        TO=59
        TOT=61
        TOK=63
        source $1
        pushd $DIR/bin/$TOOLDIR > /dev/null
        /usr/bin/time -f "\nTook %es" timeout -k $TOK $TOT $TOOL
        popd > /dev/null
\end{verbatim}
The script uses three different timeouts: \texttt{TO} is the timeout
passed to the tool itself if supported, while after \texttt{TOT} and
\texttt{TOK} the signals \texttt{SIGTERM} and \texttt{SIGKILL} are sent
to the tool in case it did not terminate on its own volition. When
multiple tools are selected, \textsf{CoCoWeb} runs them sequentially, in
order to avoid interference.

\section{Possible Extensions}
\label{sec:extensions}

We conclude this note with some ideas for future extensions of
the functionality of \textsf{CoCoWeb}.
\begin{itemize}
\item
By analyzing the format of the input problem, it is often possible to
determine the category to which the problem belongs. This information can
then be used to restrict tool selection accordingly.
\item
Adding support for other competition categories like ground confluence
and unique normal forms is an obvious extension. This, however, limits
the possibilities to restrict tool selection mentioned in the previous
item.
\item
Selected tools are executed sequentially and so if many tools are
selected, the 60 seconds time limit per tool may result in an
unacceptably slow response. In that case a timer option is a welcome
feature.
\end{itemize}

\bibliographystyle{plain}
\bibliography{references}

\end{document}